\documentstyle[aps]{revtex}

\begin{document}
\author{Gui-Fang Dang$^1$, Li-Ping Deng$^1$, and Kaige Wang$^{2,1}\thanks{%
Corresponding author: wangkg@bnu.edu.cn}$}
\address{1. Department of Physics, Applied Optics Beijing Area Major\\
Laboratory,\\
Beijing Normal University, Beijing 100875, China\\
2. CCAST (World Laboratory), P. O. Box 8730, Beijing 100080}
\title{Identifying two-photon high-dimensional entanglement in transverse patterns}
\maketitle

\begin{abstract}
We propose a scheme to explore two-photon high-dimensional entanglement
associated with a transverse pattern by means of two-photon interference in
a beamsplitter. We find that the topological symmetry of the angular
spectrum of the two-photon state governs the nature of the two-photon
interference. We prove that the anti-coalescence interference is the
signature of two-photon entanglement. On the basis of this feature, we
propose a special Mach-Zehnder interferometer incorporated with two spiral
phase plates which can change the interference from a coalescence to an
anti-coalescence type only for a two-photon entangled state. The scheme is
simple and straightforward compared with the test for a Bell inequality.

PACS number(s): 03.65.Ud, 42.50.Dv, 42.50.St
\end{abstract}

Quantum entanglement in a high-dimensional Hilbert space have potential
applications in quantum cryptography with higher alphabets and in quantum
communication with increased information flux. Recently, studies on the
conservation and entanglement of orbital angular momentum (OAM) of photons
generated by spontaneous parametric down-conversion (SPDC) have drawn much
attention\cite{mai}-\cite{pad}. This provides the possibility of realizing
photon entanglement in higher dimensions. To demonstrate the ultimate
evidence of OAM entanglement, a generalized Bell inequality should be taken
into account\cite{zei,wo}. In this work we propose a simple scheme to
identify the two-photon entanglement associated with transverse patterns by
means of two-photon interference in a beamsplitter. The two-photon
interference in a beamsplitter, first reported experimentally in Refs.\cite
{mandel,shih1}, is regarded as an effective way to test quantum nonlocality.
In Refs.\cite{wang1,wang2} the concept was put forward that the topological
symmetry of the probability-amplitude spectrum of a two-photon wavepacket in
the frequency domain determines the nature of the two-photon interference,
which may appear as coalescence interference (CI) or anti-coalescence
interference (ACI) according to the decrease or increase of the coincidence
probability in the absence of interference, respectively. The key feature is
that the ACI effect is the signature of two-photon entanglement. At the same
time, Walborn et al\cite{wal} showed that in the SPDC process the spatial
symmetry of the pump beam is related to the interference behavior of two
down-converted photons. Not limited to the SPDC case, we survey this issue
for a general two-photon state associated with transverse patterns. We prove
that ACI is evidence of two-photon entanglement. To reveal the two-photon
entanglement, we propose a special Mach-Zehnder interferometer incorporated
with two spiral phase plates (SPP), which may change the topological
symmetry of a two-photon entangled angular spectrum. This scheme can thus be
utilized to explore two-photon high-dimensional entanglement in a transverse
pattern.

To begin with, we consider a 50/50 lossless beamsplitter which satisfies a
transform relation between two input beams $a_{j}(q_{x},q_{y})$ and two
output beams $b_{j}(q_{x},q_{y})$%
\begin{eqnarray}
b_{1}(q_{x},q_{y}) &=&(1/\sqrt{2})[a_{1}(q_{x},q_{y})e^{i\varphi _{\tau
}}+a_{2}(q_{x},-q_{y})e^{i\varphi _{\rho }}],  \label{1} \\
b_{2}(q_{x},q_{y}) &=&(1/\sqrt{2})[-a_{1}(q_{x},-q_{y})e^{-i\varphi _{\rho
}}+a_{2}(q_{x},q_{y})e^{-i\varphi _{\tau }}],  \nonumber
\end{eqnarray}
where $\varphi _{\tau }$ and $\varphi _{\rho }$ are two phases allowed in
the unitary transform, and $q_{x}$ and $q_{y}$ are two orthogonally
transverse wavevectors defined in right-handed coordinates, as shown in Fig.
1. The negative sign before component $q_{y}$ refers to reflection of the
beam if $q_{y}$ is in the incident plane\cite{wal}. According to Ref.\cite
{wang1}, for an input two-photon state $|\Phi \rangle _{in}=F(a_{1}^{\dagger
},a_{2}^{\dagger })|0\rangle $, the corresponding output state is obtained
by $|\Phi \rangle _{out}=F(\overline{b}_{1}^{\dagger },\overline{b}%
_{2}^{\dagger })|0\rangle $, where $\overline{b}_{1}^{\dagger }$ and $%
\overline{b}_{2}^{\dagger }$ are written as
\begin{eqnarray}
\overline{b}_{1}^{\dagger }(q_{x},q_{y}) &=&(1/\sqrt{2})[a_{1}^{\dagger
}(q_{x},q_{y})e^{i\varphi _{\tau }}-a_{2}^{\dagger
}(q_{x},-q_{y})e^{-i\varphi _{\rho }}],  \label{2} \\
\overline{b}_{2}^{\dagger }(q_{x},q_{y}) &=&(1/\sqrt{2})[a_{1}^{\dagger
}(q_{x},-q_{y})e^{i\varphi _{\rho }}+a_{2}^{\dagger
}(q_{x},q_{y})e^{-i\varphi _{\tau }}].  \nonumber
\end{eqnarray}

For simplicity, we assume that the two photons are monochromatic and
degenerate in both polarization and frequency, but propagate in different
directions. A general two-photon state associated with the transverse field
is described as
\begin{equation}
|\Phi \rangle _{in}=\int d{\bf q}_1d{\bf q}_2\Phi ({\bf q}_1,{\bf q}%
_2)a_1^{\dagger }({\bf q}_1)a_2^{\dagger }({\bf q}_2)|0\rangle =\int d{\bf x}%
_1d{\bf x}_2\Phi ({\bf x}_1,{\bf x}_2)a_1^{\dagger }({\bf x}_1)a_2^{\dagger
}({\bf x}_2)|0\rangle ,  \label{3}
\end{equation}
where ${\bf q}_j=(q_{xj},q_{yj})$ is the transverse wavevector for photon $j$%
, and $\Phi ({\bf q}_1,{\bf q}_2)$ is the two-photon angular spectrum. The
state can also be expressed with the transverse positions by using Fourier
transformations $\Phi ({\bf x}_1,{\bf x}_2)={\cal F}^{-1}[\Phi ({\bf q}_1,%
{\bf q}_2)]\equiv (2\pi )^{-2}\int d{\bf q}_1d{\bf q}_2\Phi ({\bf q}_1,{\bf q%
}_2)\exp [i({\bf x}_1\cdot {\bf q}_1+{\bf x}_2\cdot {\bf q}_2)]$ and $a_j(%
{\bf x}_j)={\cal F}^{-1}[a_j({\bf q}_j)]$. Consider state (\ref{3}) as the
input of the beamsplitter, then the outgoing state is obtained to be
\begin{eqnarray}
|\Phi \rangle _{out} &=&(1/2)\int d{\bf q}_1d{\bf q}_2[\Phi
(q_{x1},q_{y1},q_{x2},-q_{y2})e^{i\varphi }a_1^{\dagger }({\bf q}%
_1)a_1^{\dagger }({\bf q}_2)-\Phi (q_{x1},-q_{y1},q_{x2},q_{y2})e^{-i\varphi
}a_2^{\dagger }({\bf q}_1)a_2^{\dagger }({\bf q}_2)]|0\rangle   \label{4} \\
&&+(1/2)\int d{\bf q}_1d{\bf q}_2[\Phi (q_{x1},q_{y1},q_{x2},q_{y2})-\Phi
(q_{x2},-q_{y2},q_{x1},-q_{y1})]a_1^{\dagger }({\bf q}_1)a_2^{\dagger }({\bf %
q}_2)]|0\rangle ,  \nonumber
\end{eqnarray}
where $\varphi =\varphi _\tau +\varphi _\rho $. The coincidence probability
in the two outgoing arms is evaluated as
\begin{equation}
P_c=(1/2)[1-\int d{\bf q}_1d{\bf q}_2\Phi (q_{x1},q_{y1},q_{x2},q_{y2})\Phi
^{*}(q_{x2},-q_{y2},q_{x1},-q_{y1})].  \label{5}
\end{equation}
When the two-photon angular spectrum satisfies the symmetric condition, $%
\Phi (q_{x1},q_{y1},q_{x2},q_{y2})=\Phi (q_{x2},-q_{y2},q_{x1},-q_{y1})$,
the coincidence probability is null and perfect CI occurs. Contrarily, when
the spectrum satisfies the anti-symmetric condition $\Phi
(q_{x1},q_{y1},q_{x2},q_{y2})=-\Phi (q_{x2},-q_{y2},q_{x1},-q_{y1})$, the
coincidence probability is unity and perfect ACI occurs. In the latter case,
we can prove that $|\Phi \rangle _{out}=|\Phi \rangle _{in}$, that is, the
beamsplitter becomes transparent due to the quantum interference\cite{wang1}%
. In polar coordinates ($q,\theta $), the conditions are written as $\Phi
(q_1,q_2,\theta _1,\theta _2)=\pm \Phi (q_2,q_1,2\pi -\theta _2,2\pi -\theta
_1)$.

For a non-entangled two-photon state, $\Phi ({\bf q}_{1},{\bf q}_{2})=\Phi
_{1}({\bf q}_{1})\Phi _{2}({\bf q}_{2})$, we calculate the integral in Eq. (%
\ref{5}) $\int d{\bf q}_{1}d{\bf q}_{2}\Phi
(q_{x1},q_{y1},q_{x2},q_{y2})\Phi ^{\ast
}(q_{x2},-q_{y2},q_{x1},-q_{y1})=|\int dq_{x}dq_{y}\Phi
_{1}(q_{x},q_{y})\Phi _{2}^{\ast }(q_{x},-q_{y})|^{2}\geq 0$, so that ACI
never happens ($P_{c}\leq 1/2$). In other words, we can identify two-photon
entanglement by the condition $P_{c}>1/2$. We note that the above discussion
is also valid in the position space after replacing ${\bf q}_{j}$ by ${\bf x}%
_{j}$.

We now show some examples. The first example seems to contradict our common
knowledge of two-photon interference. Two independent degenerate photons
with the same OAM $|l,l\rangle $ do not interfere in a beamsplitter ($%
P_c=1/2 $), whereas two photons with opposite OAM $|l,-l\rangle $ show
perfect CI ($P_c=0$). This is because the transmission does not change a
photon's OAM while the reflection results in an opposite OAM. Then we
consider a set of Bell states involving the photon's OAM
\begin{mathletters}
\label{6}
\begin{eqnarray}
|\Psi ^{\pm }\rangle &=&(1/\sqrt{2})(|l,l\rangle \pm |-l,-l\rangle )=\int d%
{\bf q}_1d{\bf q}_2R(q_1)R(q_2)(e^{il(\theta _1+\theta _2)}\pm e^{-il(\theta
_1+\theta _2)})a_1^{\dagger }({\bf q}_1)a_2^{\dagger }({\bf q}_2)|0\rangle ,
\label{6a} \\
|\Phi ^{\pm }\rangle &=&(1/\sqrt{2})(|l,-l\rangle \pm |-l,l\rangle )=\int d%
{\bf q}_1d{\bf q}_2R(q_1)R(q_2)(e^{il(\theta _1-\theta _2)}\pm e^{-il(\theta
_1-\theta _2)})a_1^{\dagger }({\bf q}_1)a_2^{\dagger }({\bf q}_2)|0\rangle .
\label{6b}
\end{eqnarray}
It is readily seen that $|\Psi ^{+}\rangle $ and $|\Phi ^{\pm }\rangle $
satisfy the symmetric condition while $|\Psi ^{-}\rangle $ satisfies the
anti-symmetric one. This feature is similar to the Bell basis associated
with polarization and frequency\cite{wang3}. However, for the OAM Bell
bases, two of them can be gained from the other two through the reflection
of the two photons, i.e. $|\Psi ^{\pm }\rangle \leftrightarrow $ $|\Phi
^{\pm }\rangle $.

In the SPDC process, the signal and idler photon pair can be described as a
two-photon state
\end{mathletters}
\begin{equation}
|\Phi \rangle =\frac{1}{\pi }\sqrt{\frac{2L}{k_{p}}}\int d{\bf q}_{1}d{\bf q}%
_{2}\upsilon ({\bf q}_{1}+{\bf q}_{2})\text{sinc}[L|{\bf q}_{1}-{\bf q}%
_{2}|^{2}/(4k_{p})]a_{1}^{\dagger }({\bf q}_{1})a_{2}^{\dagger }({\bf q}%
_{2})|0\rangle ,  \label{7}
\end{equation}
where $\upsilon ({\bf q})$ is the angular spectrum of the pump beam with the
wavenumber $k_{p}$ and $L$ is the length of the crystal. In Eq. (\ref{7}),
the sinc function satisfies the symmetric condition, so that the symmetry of
$\upsilon ({\bf q})$ dominates the nature of the two-photon interference.
When the pump beam is in the Gaussian fundamental mode with a beam waist $%
w_{0}$, $\upsilon ({\bf q})=G_{00}({\bf q})=(w_{0}/\sqrt{2\pi })\exp (-|{\bf %
q}|^{2}w_{0}^{2}/4),$ $\upsilon ({\bf q}_{1}+{\bf q}_{2})$ satisfies the
symmetric condition. If the mode structure of the pump beam is
Hermitian-Gaussian $HG_{mn}({\bf q})$, then symmetric and anti-symmetric
two-photon spectra are obtained for even and odd numbers of the subscript $n$%
, respectively. This result has been demonstrated experimentally in Fig. 3
of Ref. \cite{wal}, where CI and ACI occur for the pump modes $HG_{10}$ and $%
HG_{01}$, respectively, in the case of the symmetric polarization.

To explore two-photon entanglement for an angular spectrum satisfying the
symmetric condition, we propose a scheme which may change the symmetry of an
entangled two-photon spectrum without introducing any additional
entanglement. The key device in the scheme, shown in Fig. 2, is a
Mach-Zehnder interferometer (MZI) containing two identical spiral phase
plates (SPPs). The role of the latter is to produce a phase shift $\exp
(i\varsigma \theta )$ linearly distributed along the azimuthal angle $\theta
$\cite{wo}. Because the two SPPs are the same, there is coherent
superposition of the phases $\exp (i\varsigma \theta )$ and $\exp
(-i\varsigma \theta ),$ due to transmission and reflection, respectively, in
each outgoing arm of the MZI. An adjustable phase shifter (PS) is inserted
into the path of beam $a_1$. Note that the two beams are reflected twice by
the mirrors before the last beamsplitter BS, so that the spectrum function
is invariant. An angular spectrum $\Phi ({\bf q}_1,{\bf q}_2)$ of a
two-photon state is defined at the source. In the paraxial approximation, a
beam propagating a distance $z$ will introduce a phase $\exp [ikz-iq^2z/(2k)]
$. While beam $a_2$ freely propagates a distance $z_2$, beam $a_1$ is
divided into two beams by beamsplitter BS1 and travels a distance $z_1$ to
the SPP, thus the two-photon state can be written as
\begin{equation}
|\Phi \rangle _1=(1/\sqrt{2})e^{ik(z_1+z_2)}\int d{\bf q}_1d{\bf q}%
_2e^{-i(q_1^2z_1+q_2^2z_2)/(2k)}[\Phi ({\bf q}_1,{\bf q}_2)e^{i\varphi
_{1\tau }}a_1^{\dagger }({\bf q}_1)-\Phi (q_{x1},-q_{y1},{\bf q}%
_2)e^{-i\varphi _{1\rho }}a_3^{\dagger }({\bf q}_1)]a_2^{\dagger }({\bf q}%
_2)|0\rangle ,  \label{8}
\end{equation}
where $\varphi _{j\tau }$ and $\varphi _{j\rho }$ are the phases associated
with BS-$j$ ($j$=1,2). The state can also be expressed with the position
variables. Considering the phases $\exp (i\varsigma \theta )$ and $\exp
(i\varphi )$ introduced by the SPPs and the phase shifter, respectively, we
obtain
\[
|\Phi \rangle _2=(1/\sqrt{2})\int d{\bf x}_1d{\bf x}_2[\Psi ({\bf x}_1,{\bf x%
}_2,z_1,z_2)e^{i(\varphi _{1\tau }+\varphi )}a_1^{\dagger }({\bf x}_1)-\Psi
(x_1,-y_1,{\bf x}_2,z_1,z_2)e^{-i\varphi _{1\rho }}a_3^{\dagger }({\bf x}%
_1)]e^{i\varsigma \theta _1}a_2^{\dagger }({\bf x}_2)|0\rangle ,
\]
where $\Psi ({\bf x}_1,{\bf x}_2,z_1,z_2)={\cal F}^{-1}[\Phi ({\bf q}_1,{\bf %
q}_2)e^{ik(z_1+z_2)-i(q_1^2z_1+q_2^2z_2)/(2k)}]$. After passing through BS2,
the two-photon state as it arrives at the last BS is given by
\begin{equation}
|\Phi \rangle _3=ie^{i(\varsigma \pi +\alpha _{-})}\int d{\bf x}_1d{\bf x}%
_2\Psi ({\bf x}_1,{\bf x}_2,z_1,z_2)\sin [\varsigma (\theta _1-\pi )+\alpha
_{+}]a_1^{\dagger }({\bf x}_1)a_2^{\dagger }({\bf x}_2)|0\rangle ,  \label{9}
\end{equation}
where $\alpha _{\pm }=(1/2)[\varphi +\varphi _{1\tau }+\varphi _{2\tau }\pm
(\varphi _{1\rho }-\varphi _{2\rho })]$. Beam $a_3$ has been omitted here
since it is not involved in the interference in the last BS. For simplicity,
we assume that the SPPs are close to the last BS and hence the propagation
distance between the SPPs and the last BS can be neglected\cite{note}.

Equation (\ref{9}) is compatible with that of the similar scheme for
two-photon entanglement in the frequency domain, which was experimentally
implemented\cite{chiao,shih2} and later theoretically explained\cite{wang3}.
As an example, we consider a two-photon entangled state generated by SPDC
with a thin crystal where the sinc function is not taken into account. When
the Gaussian mode $G_{00}$ pumps the crystal, we obtain
\begin{equation}
\Psi ({\bf x}_1,{\bf x}_2,z,z)=A\exp \left\{ -\frac{|{\bf x}_1+{\bf x}_2|^2}{%
4w^2(z)}+i\frac{k_p}4\left[ \frac{z_0^2|{\bf x}_1-{\bf x}_2|^2}{2z^2R(z)}+%
\frac{|{\bf x}_1|^2+|{\bf x}_2|^2}{R(z)}\right] \right\} ,  \label{10}
\end{equation}
where $k_p=2k$ is the wavenumber of the pump beam, $z_0=k_pw_0^2/2$ the
Rayleigh length, $w(z)=w_0(1+z^2/z_0^2)^{1/2}$ the spot size and $%
R(z)=(z^2+z_0^2)/z$ the radius of curvature of the Gaussian beam. According
to Eq. (\ref{5}), the phase part in Eq. (\ref{10}) does not contribute to
the coincidence probability. In Fig. 3a, we show the numerical result of the
coincidence probability as a function of the parameter $\varsigma $ using
Eqs. (\ref{5}), (\ref{9}) and (\ref{10}). For $\alpha _{+}=0$ ($\alpha
_{+}=\pi /2$), the odd and even (even and odd) numbers $\varsigma $
contribute to maximum CI and ACI effects, respectively. This result can be
explained by the symmetry of the wavefunction. When the aperture of the
optical system is much larger than the spot size $w(z)$, the Gaussian
function in Eq. (\ref{10}) is closer to $\delta ({\bf x}_1+{\bf x}_2)$,
resulting in $\theta _1=\theta _2+\pi $. Relying on the strong correlation
between $\theta _1$ and $\theta _2$ of the two photons, the function $\sin
[\varsigma (\theta _1-\pi )]$ can satisfy either the symmetric or
anti-symmetric conditions provided $\varsigma $ is taken as odd or even,
respectively. Figure 3b shows the coincidence probability versus the phase $%
\alpha _{+}$ for $\varsigma =1$ and $2$. Experimentally, the reference for
the coincidence probability can be acquired by setting a path difference for
the two photons so that interference does not occur in the beamsplitter. The
coincidence probability larger than this reference witnesses two-photon
entanglement.

In summary, we have demonstrated that the nature of two-photon interference
in a beamsplitter is related to the topological symmetry of the two-photon
angular spectrum. The ACI effect is the signature of two-photon
entanglement. On the basis of this feature, we have proposed a scheme which
can identify two-photon high-dimensional entanglement in transverse patterns
by means of a special Mach-Zehnder interferometer.

The authors thank L. A. Wu for helpful discussions. This research was
supported by the National Fundamental Research Program of China Project No.
2001CB309310, and the National Natural Science Foundation of China, Project
No. 60278021 and No. 10074008.

Figure Captions

Fig. 1. Sketch of two-photon interference in a beamsplitter. Right handed
coordinates are used for each beam.

Fig. 2. A scheme to explore two-photon entanglement in transverse patterns.
Beam $a_{1}$ passes through a Mach-Zehnder interferometer (MZI) consisting
of two beamsplitters (BS1 and BS2) and two mirrors. Two identical spiral
phase plates (SPP) are put in each path of the MZI and an adjustable phase
shifter (PS) is in the path of beam $a_{1}$. After MZI, beams $a_{1}$ and $%
a_{2}$ interfere in the last beamsplitter (BS) and coincidence measurement
is performed for the two outgoing beams.

Fig. 3. (a) Coincidence probability as a function of the SPP\ parameter $%
\varsigma $ for $\alpha _{+}=0$ and $\pi /2$; (b) Coincidence probability as
a function of the phase $\alpha _{+}$ for the SPP\ parameter $\varsigma $ $%
=1 $ and $2$. In the numerical simulation, the aperture of the optical
system is 40 times larger than the spot size $w(z)$.


\begin{references}
\bibitem{note}  The further theoretical analysis shows that the phase
introduced in the propagation between the SPPs and the last BS does not
affect the result if the two photons travel the same path.
\end{references}
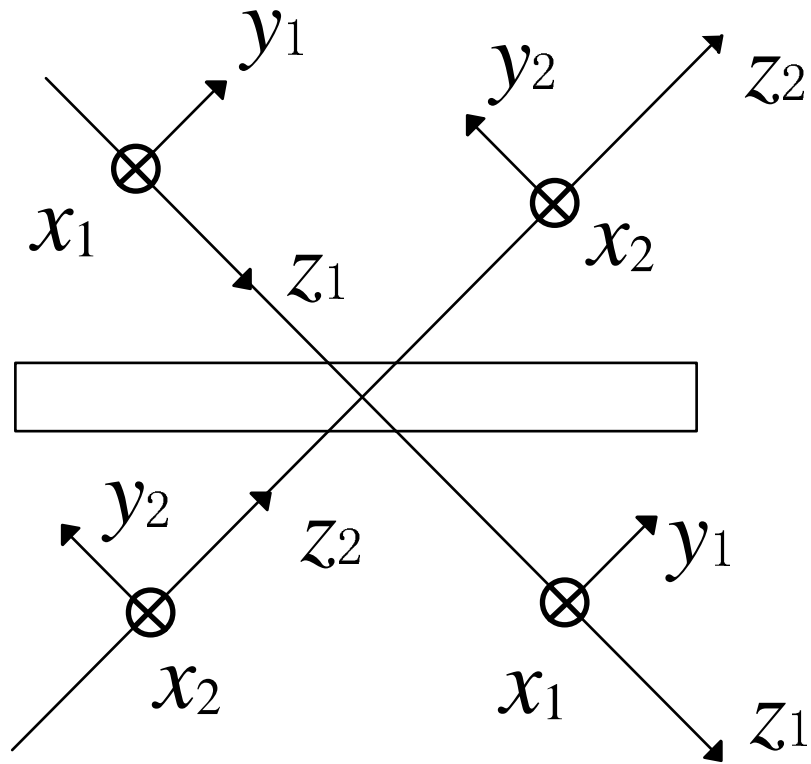
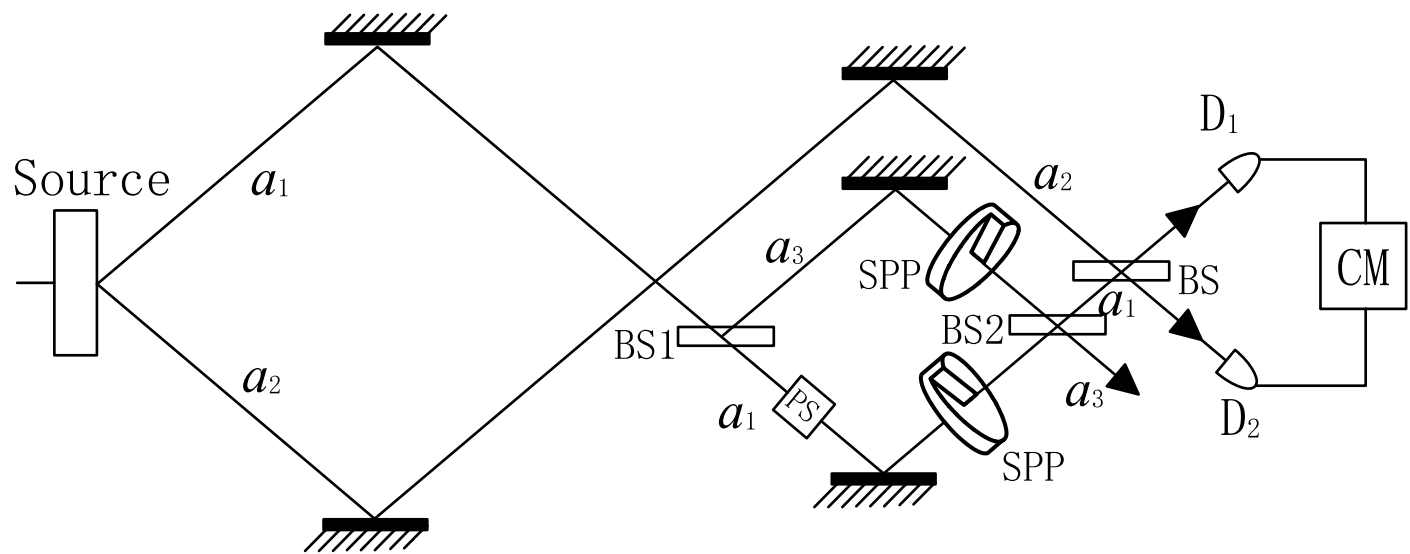
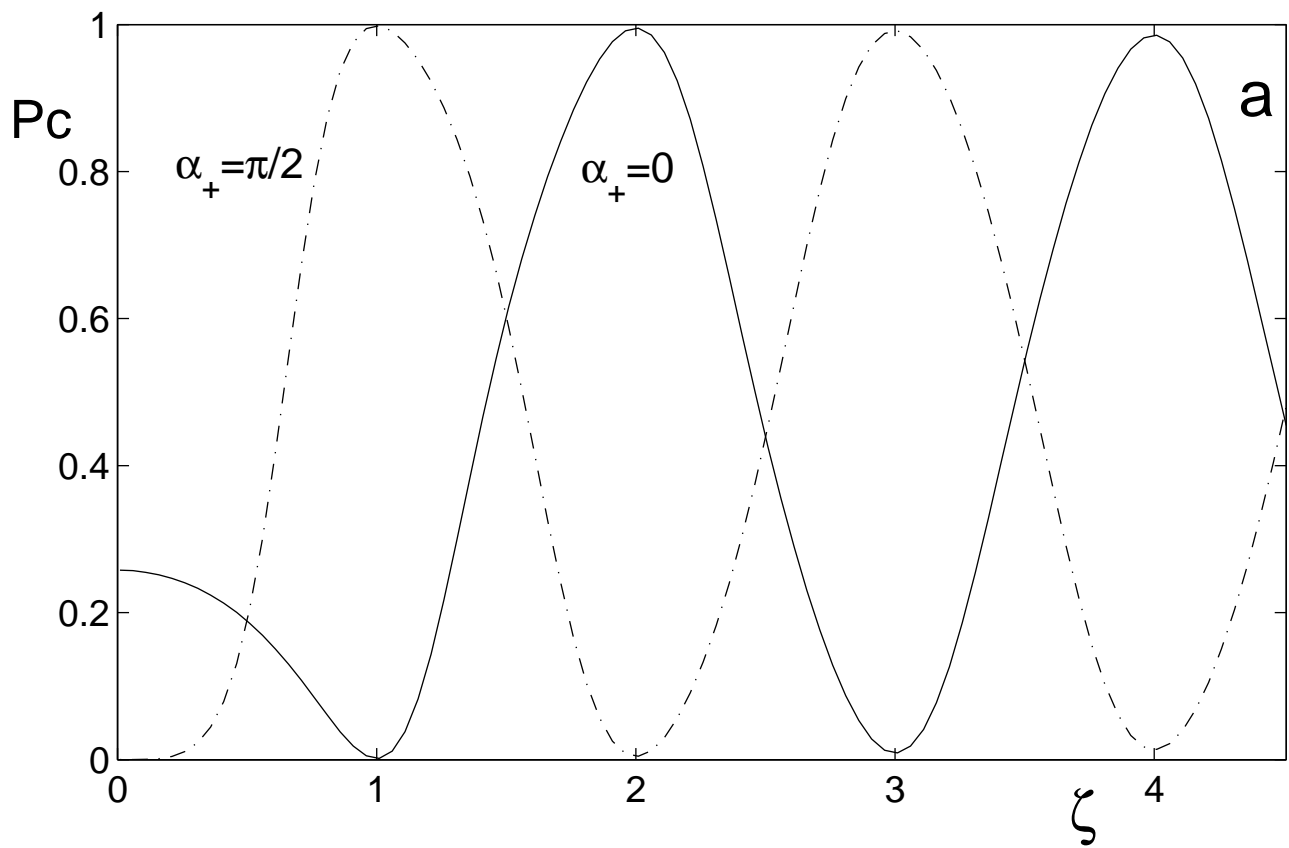
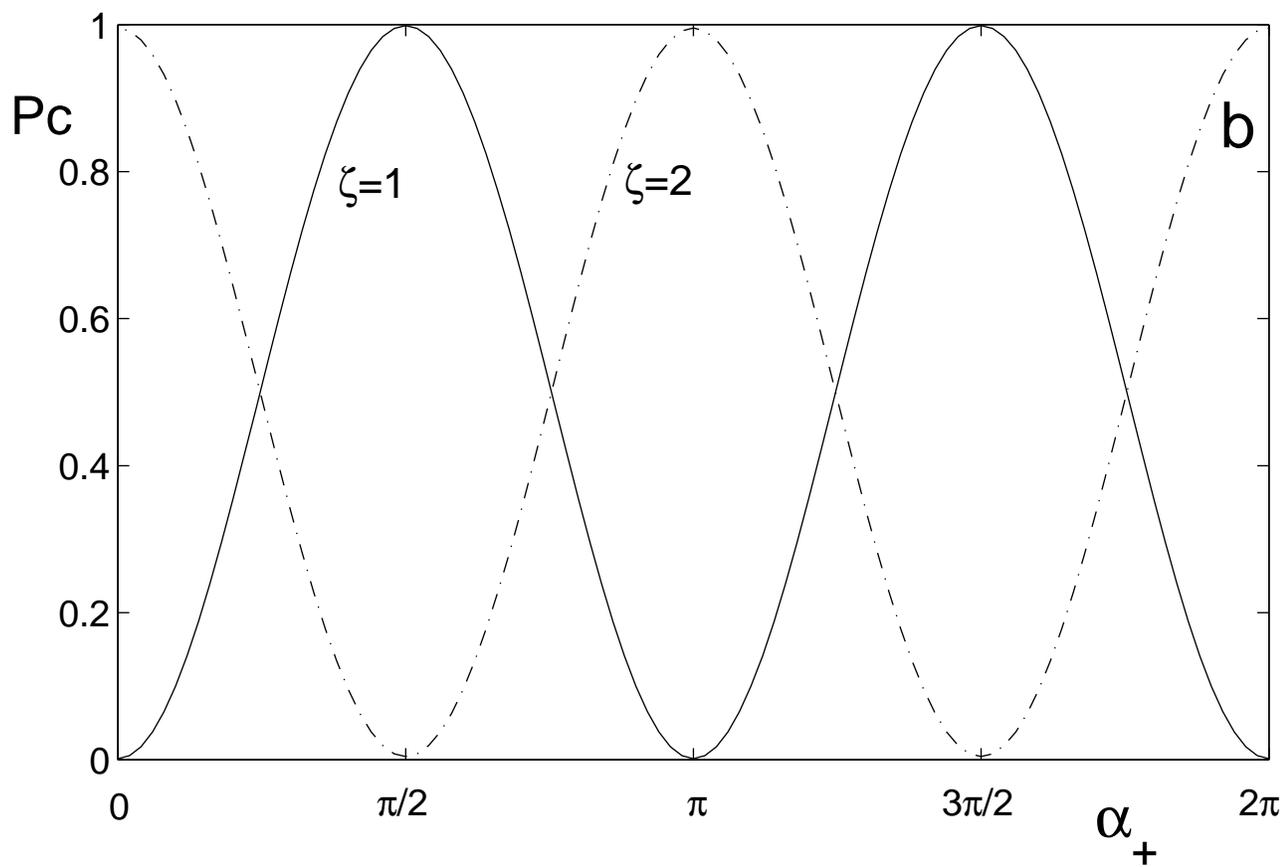
\end{document}